\documentclass{ws-p9-75x6-50}
\newcommand {\beqa}{\begin{eqnarray}}
\newcommand {\eeqa}{\end{eqnarray}}
\newcommand {\beq}{\begin{equation}}
\newcommand {\eeq}{\end{equation}}
\newcommand {\qslash}{q\!\!\!/}
\newcommand {\muslash}{\mu\!\!\!/}

\begin{document}

\title{Color Superconductivity}

\author{Thomas Sch\"afer}

\address{TRIUMF, 4004 Wesbrook Mall\\
Vancouver, BC, Canada, V6T2A3}


\maketitle

\abstracts{We discuss recent results on color superconductivity
in QCD at large chemical potential.}

\section{Introduction}

  The phase structure of matter at non-zero baryon density
has recently attracted a great deal of interest. The behavior 
of hadronic matter in this regime is of interest in 
connection with the structure of compact astrophysical objects 
and the physics of heavy ion collisions in the regime of 
maximum baryon density. In addition to that, we are 
addressing a fundamental question about QCD, namely, what
is the ultimate fate of hadronic matter as we keep 
compressing it to higher and higher densities. And finally,
understanding the structure of matter at finite density is
part of our quest to understand the full richness of the 
QCD phase diagram as we vary the temperature, the density,
and the number of quark flavors and their masses. 

  At very high density the natural degrees of freedom are
quasiparticles and holes in the vicinity of the Fermi
surface. Since the Fermi momentum is large, asymptotic freedom
implies that the interaction between quasiparticles is weak.
In QCD, because of the presence of unscreened long range gauge 
forces, this is not quite true. Nevertheless, we believe 
that this fact does not essentially modify the argument. 
We know from the theory of superconductivity that the Fermi 
surface is unstable in the presence of even an arbitrarily 
weak attractive interaction. At very large density, the 
attraction is provided by one-gluon exchange between 
quarks in a color anti-symmetric $\bar 3$ state. QCD at high 
density is therefore expected to behave as a color superconductor 
\cite{Frau_78,Bar_77,BL_84}.

  Color superconductivity is described by an order 
parameter of the general form 
\be
\label{csc}
\phi = \langle \psi^TC\Gamma_D\lambda_C\tau_F\psi\rangle.
\ee
Here, $C$ is the charge conjugation matrix, and $\Gamma_D,
\lambda_C,\tau_F$ are Dirac, color, and flavor matrices. 
Except in the case of only two colors, the order parameter
cannot be a color singlet. Color superconductivity is 
therefore characterized by the breakdown of color gauge 
invariance. As usual, this statement has to be interpreted 
with care. Local gauge invariance cannot really be broken 
\cite{Eli_75}. Nevertheless, we can fix the gauge and
study, in weak coupling perturbation theory, the 
implications of a non-zero vacuum expectation value 
of the type (\ref{csc}). The most important gauge invariant 
consequence of color superconductivity is the appearance
of a mass gap, through the Anderson-Higgs phenomenon. 
The formation of a mass gap is of course also characteristic
of a confined phase. Indeed, it is known that in general 
Higgs and confined phases are continuously connected \cite{FS_79}.

 In addition to that, color superconductivity can lead to 
the breakdown of global symmetries. We shall see that in 
some cases there is a gauge invariant order parameter 
for the breaking of the $U(1)$ of baryon number. This 
corresponds to true superfluidity and the appearance of a 
massless phonon. We shall also find that for $N_f>2$ color 
superconductivity leads to chiral symmetry breaking.

\section{The renormalization group}
\label{sec_rge}

  In order to assess the relative importance of different
instabilities of the quark Fermi liquid we employ the
renormalization group methods described in \cite{Pol_92,Sha_95}.
The idea is quite simple. The low energy excitations of a
degenerate Fermi systems are quasiparticles and holes in
the vicinity of the Fermi surface, described by an effective 
action of the form
\be
 S_{eff} = \int dt\, d^3p\; \psi^\dagger \left( i\partial_t
 - (\epsilon(p)-\epsilon_F) \right)\psi + S_{int}.
\ee
Quasiparticle interactions are represented by a tower of $2n$ 
fermion operators. We can analyze the general structure of the 
interactions by studying the evolution of the corresponding 
operators as we successively integrate out modes closer and
closer to the Fermi surface. The main result of this analysis
is that, in general, four fermion, six fermion, and higher 
order interactions are suppressed as we approach the Fermi 
surface. This fixed point corresponds to Landau liquid theory. 
The only exceptions are the well known instabilities in the 
particle-particle and particle-hole channels. In the 
particle-particle channel the dominant operators correspond
to two particles from opposite corners of the Fermi surface 
$(p_F,-p_F)$ scattering into $(p'_F,-p'_F)$. This kind of 
scattering leads to a logarithmic growth of the coupling 
constant as we approach the Fermi surface, and the BCS 
instability. 

  In QCD, a complete set of four fermion operators is given by
\beqa
\label{nf3_ops}
O^0_{LL} &=& (\bar\psi_L\gamma_0\psi_L)^2, \hspace{1cm}
O^0_{LR} \;=\; (\bar\psi_L\gamma_0\psi_L)(\bar\psi_R\gamma_0\psi_R) \\
O^i_{LL} &=& (\bar\psi_L\gamma_i\psi_L)^2,  \hspace{1cm}
O^i_{LR} \;=\; (\bar\psi_L\vec\gamma\psi_L)
       (\bar\psi_R\vec\gamma\psi_R) . \nonumber
\eeqa
Each of these operators comes in two color structures, 
for example color symmetric and color anti-symmetric 
\beqa
(\bar\psi^a\psi^b)(\bar\psi^c\psi^d)
  \left(\delta_{ab}\delta_{cd}\pm \delta_{ad}\delta_{bc}\right).
\eeqa
Nothing essentially new emerges from considering superficially 
different isospin structures, or different Dirac matrices. All 
such structures can be reduced to linear combinations of the 
basic ones (\ref{nf3_ops}), or their parity conjugates, by 
Fierz rearrangements. In total, we have to consider eight 
operators. 

  We should emphasize that we have restricted ourselves to 
massless QCD, and local operators invariant under the full
$SU(3)_L\times SU(3)_R\times U(1)_A$ chiral symmetry. The
extension to operators that break the anomalous $U(1)_A$
symmetry is discussed in \cite{SW_98,EHS_98}. Also, QCD
at high density contains long range interactions due to
unscreened magnetic gluon exchanges. The inclusion of long
range forces is discussed in \cite{Son_98,HS_99} and in 
section \ref{sec_oge} below.

  The operators (\ref{nf3_ops}) are renormalized by quark-quark 
scattering in the vicinity of the Fermi surface. This means that both 
incoming and outgoing quarks have momenta $\vec p_1, \vec p_2\simeq
\pm \vec p$ and $\vec p_3, \vec p_4\simeq\pm \vec q$ with $|\vec p|, 
|\vec q| \simeq p_F$. We can take the external frequency to be zero. 
A graph with vertices $\Gamma_1$ and $\Gamma_2$ then gives 
\be
\label{loop}
G_1 G_2 I\; (\Gamma_1)_{i'i}(\Gamma_1)_{k'k}
  \left[ -(\gamma_0)_{ij}(\gamma_0)_{kl}+\frac{1}{3}
          (\vec\gamma)_{ij}(\vec\gamma)_{kl}\right]
 (\Gamma_2)_{jj'}(\Gamma_2)_{ll'}
\ee
with $I=\frac{i}{8\pi^2}\mu^2\log(\Lambda_{IR}/\Lambda_{UV})$. 
Here $[\Lambda_{IR},\Lambda_{UV}]$ is the range of momenta that
was integrated out. We will denote the density of states on the 
Fermi surface by $N=\mu^2/(2\pi^2)$ and the logarithm of the scale
$t=\log(\Lambda_{IR}/\Lambda_{UV})$. The renormalization group 
does not mix $LL$ and $LR$ operators, and it also does not mix 
different color structures. The evolution equations are
\beqa 
\label{nf2_evol}
\frac{d(G^{LL}_0+G^{LL}_i)}{dt} &=& -\frac{N}{3}
   (G^{LL}_0+G^{LL}_i)^2, \\
\frac{d(G^{LL}_0-3G^{LL}_i)}{dt} &=& -N
   (G^{LL}_0-3G^{LL}_i)^2, \\
\frac{d(G^{LR}_0+3G^{LR}_i)}{dt} &=& 0,\\
\frac{d(G^{LR}_0-G^{LR}_i)}{dt} &=& -\frac{2N}{3}
   (G^{LR}_0-G^{LR}_i)^2.
\eeqa
In this basis the evolution equations are already diagonal. The coupling
$G_1=G^{LL}_0+G^{LL}_i$ evolves as 
\be
 G_1(t) = \frac{1}{1+(N/3)G_1(0)t}
\ee
with analogous results for the other operators. Note that the evolution
starts at $t=0$ and moves towards the Fermi surface $t\to-\infty$.
If the coupling is attractive at the matching scale, $G_1(0)>0$, it 
will grow during the evolution, and reach a Landau pole at $t_c=3/(N
G_1(0))$. The corresponding energy scale is 
\be 
\Lambda_{IR} = \Lambda_{UV} \exp\left(-\frac{3}{NG_1(0)}\right).
\ee
This is the standard BCS result. The location of the pole is controlled 
by the initial value of the coupling and the coefficient in the evolution 
equation. If the initial coupling is negative, the coupling decreases 
during the evolution. The second operator in (\ref{nf2_evol}) has the 
largest coefficient and will reach the Landau pole first, unless the 
initial value is very small or negative. In that case, subdominant 
operators may determine the pairing.  

  The particular form of the operators that diagonalize the 
evolution equations can be made more transparent using a Fierz 
transformation. We find
\beqa
O_{dom} &=&   2(\psi_LC \psi_L)(\bar\psi_L C\bar\psi_L),\\
O_{sub,1} &=& \frac{1}{3}(\psi_L C\vec\gamma\psi_R)
 (\bar\psi_RC\vec\gamma\psi_L) + \ldots ,\\
O_{sub,2} &=& \frac{4}{3}(\psi_L C\vec\Sigma\psi_L)
 (\bar\psi_LC\vec\Sigma \bar\psi_L) ,\\
O_{mar}   &=&  \frac{1}{2}(\psi_LC\gamma_0\psi_R)
 (\bar\psi_RC\gamma_0\psi_L) + \ldots ,
\eeqa
where we have ordered the operators according to the
size of the coefficient in the evolution equations. We 
find that the dominant operator corresponds to pairing 
in the scalar diquark channel, while the subdominant 
operators contain vector diquarks. Note that from the 
evolution equation alone we cannot decide what the preferred 
color channel is. To decide this question, we must invoke 
the fact that ``reasonable'' interactions, like one gluon 
exchange, are attractive in the color anti-symmetric
repulsive in the color symmetric channel. This can be
seen from the identity
\be
(\vec\lambda)^{ab}(\vec\lambda)^{cd}=
 \frac{2}{3} (\delta^{ab}\delta^{cd}+\delta^{ad}\delta^{bc})
-\frac{4}{3} (\delta^{ab}\delta^{cd}-\delta^{ad}\delta^{bc}).
\ee
If the color wave function is anti-symmetric, the Pauli
exclusion principle fixes the isospin wave function to 
be anti-symmetric as well. The dominant operator does not 
distinguish between scalar and pseudoscalar diquarks.   
This degeneracy is lifted by operators that break the
axial baryon number symmetry. These operators are 
associated with instantons. As we will discuss in more
detail in section \ref{sec_inst}, instantons favor
scalar diquarks over pseudoscalar diquarks.

\section{The effective potential}
\label{sec_veff}

  The form of the dominant operator indicates the existence 
of potential instabilities, but does not itself indicate 
how they are resolved. In this section we wish to introduce 
a simple energy functional that captures the essential dynamics 
of QCD at high baryon density. We shall use this functional in 
order to analyze the ground state of QCD with different 
numbers of colors and flavors.

 The dominant coupling is a color and flavor ant-symmetric 
interaction of the form
\be
\label{laa}
 {\cal L} = G
 \big(\delta^{ac}\delta^{bd}-\delta^{ad}\delta^{bc}\big)
 \big(\delta_{ik}\delta_{jl}-\delta_{il}\delta_{jk}\big)
  \left(\psi^a_i C\gamma_5 \psi^b_j \right)
     \left(\bar\psi^c_k C\gamma_5 \bar\psi^d_l \right)
\ee
where $a,b,\ldots$ are color indices and $i,j\ldots$ 
are flavor indices. In the following, we shall use the 
notation ${\cal K}^{abcd}_{ijkl}$ for the color-flavor 
structure of the interaction. 

  In order to determine the structure of the ground state
we have to calculate the grand canonical potential of the system 
for different trial states. Since the interaction is
attractive in s-wave states, it seems clear that the 
dominant order parameter is an s-wave, too. We then 
only have to study the color-flavor structure of the 
primary condensate. We assume that the condensate 
takes the form
\be
\label{order}
 \langle \psi^a_i C\gamma_5\psi^b_j\rangle
 = \phi^{ab}_{ij}.
\ee
$\phi^{ab}_{ij}$ is a $N_f\times N_f$ matrix in flavor
space and a $N_c\times N_c$ matrix in color space. 

  We calculate the effective potential using the bosonization
method. For this purpose, we introduce collective fields
$\Delta^{ab}_{ij}$ and $\bar\Delta^{ab}_{ij}$ with the 
symmetries of the order parameter (\ref{order}). We add to
the fermionic action a term $G^{-1}{\cal K}^{abcd}_{ijkl}
\Delta^{ab}_{ij}\bar\Delta^{cd}_{kl}$ and integrate over the 
dummy variables $\Delta^{ab}_{ij}$ and $\bar\Delta^{ab}_{ij}$. 
We then shift the integration variables to eliminate the 
interaction term (\ref{laa}). So far, no approximations have
been made. We now assume that the collective fields are 
slowly varying, and that $\Delta^{ab}_{ij}$ can be replaced
by a constant. In this case, we can perform the integration
over the fermionic fields and determine the grand canonical 
potential as a function of the gap matrix $\Delta^{ab}_{ij}$. 

 The integration over the fermions is performed using the 
Nambu-Gorkov formalism. We introduce a two component field 
$\Psi=(\psi,\bar\psi^T)$. The inverse quark propagator takes 
the form
\beq
\label{sinv}
S^{-1}(q) = \left(\begin{array}{cc}
 \qslash+\muslash-m &  {\cal K}\cdot\bar\Delta \\
 {\cal K}\cdot\Delta  & (\qslash-\muslash+m)^T 
\end{array}\right).
\eeq
The grand canonical potential is now given by
\beq 
\Omega(\Delta) = \frac{1}{2}{\rm Tr}\left[\log(S)\right]
 + \frac{1}{G}\Delta\cdot{\cal K}\cdot\bar\Delta.
\eeq
In order to evaluate the logarithm, we have to diagonalize 
the mass matrix ${\cal M}={\cal K}\cdot \Delta$. Let us denote 
the corresponding eigenvalues by $\delta_\rho\,(\rho=1,\ldots,
N_cN_f)$. These are the physical gaps of the $N_fN_c$ fermion 
species. The ${\rm tr}\log(S)$ term has an ultra-violet divergence. 
This divergence can be regularized in a way that is consistent
with the renormalization group approach \cite{Wei_94}. We find
\beq
\label{renp}
 \Omega_{ren}(\Delta) = \sum_{\rho}\left\{ 
 -\frac{\mu^2}{4\pi^2}\delta_\rho^2
  \left(\log\left(\frac{\delta_\rho}{\xi}\right)-1\right)
 + \frac{4}{G(\xi)}\delta_\rho^2 \right\}.
\eeq
Here, $\xi$ is a renormalization scale. The grand potential
is independent of $\xi$, since the scale dependence of the
first term is canceled by the scale dependence of the coupling 
constant $G$. The coupling constant satisfies the renormalization
group equation discussed above.

\section{Superconductivity in QCD with $N_c$ colors and $N_f$
flavors}
\label{sec_nf}

 From the minima of the effective potential we can now
determine the groundstate properties in a theory with
$N_c$ colors and $N_f$ flavors. The simplest case is 
QCD with two colors and two flavors. In this case, we
have  
\be
\label{nc_2}
 \Delta^{ab}_{ij} = \Delta \epsilon^{ab}\epsilon_{ij},
\hspace{0.5cm} (N_c=N_f=2).
\ee
This order parameter is a color and flavor singlet. This
would seem to indicate that no symmetries are broken, but
this conclusion is not correct. QCD with two colors has
a particle-anti-particle (Pauli-G\"ursey) symmetry which
enlarges the flavor symmetry of the massless theory from 
$SU(2)\times SU(2)$ to $SU(4)$. Spontaneous chiral 
symmetry breaking by a non-zero quark condensate
$\langle\bar\psi\psi\rangle$ corresponds to the
symmetry breaking pattern $SU(4)\to Sp(4)$. The 
coset $SU(4)/Sp(4)=S^5$, so there are 5 Goldstone
bosons. These Goldstone bosons are the three pions,
the scalar diquark, and its anti-particle. Under the
Pauli-G\"ursey symmetry the chiral condensate 
$\langle\bar\psi\psi\rangle$ is equivalent to the
diquark condensate $\langle\epsilon^{ab}\epsilon_{ij}
\psi^a_i C\gamma_5\psi^b_j\rangle$. In the diquark
condensed phase, the five Goldstone bosons are the
three pions, the sigma, and the anti-scalar diquark. 

 Both a quark mass term and a chemical potential 
break the $SU(4)$ symmetry. If the chemical potential
is zero, but the quark mass is not, then the theory
has a non-zero chiral condensate. If the chemical 
potential is increased there is a phase transition
at a critical chemical potential $\mu_c\simeq m_\pi/2$.
For $\mu>\mu_c$, the order parameter rotates from 
the chiral condensate to the diquark condensate. 
QCD with two colors is discussed in more detail in 
\cite{RSSV_98,KST_99}.
  
  We would now like to discuss real QCD with three
colors. The case of only one flavor is special, since
in this case a color anti-symmetric condensate cannot
be a spin singlet. We therefore expect condensation 
to take place in a spin-1 channel. This problem is
of some relevance for the behavior of real QCD with 
two light and one intermediate mass flavor. For $m_s$ 
larger than some critical value, QCD has a phase with 
separate pairing in the $ud$ and $s$ sectors 
\cite{SW_99,ABR_99}. Here, we will not discuss
$N_f=1$ QCD any further. 

  In the case of two flavors, the order parameter
\beq
\label{order_2}
 \Delta^{ab}_{ij} = \Delta \epsilon^{3ab}\epsilon_{ij},
 \hspace{0.5cm} (N_c=3,\;N_f=2),
\eeq
breaks color $SU(3)\to SU(2)$. The chiral $SU(2)_L
\times SU(2)_R$ symmetry remains unbroken. At this level,
the Fermi surfaces of the up and down quarks of the third
color remain intact. Subleading interactions can generate
a gap for these states. The exact nature of this gap is 
hard to determine, even in the limit of very large chemical
potential. 

  The most interesting case is the situation for three
colors and three flavors. As shown in \cite{SW_99,ABR_99},
these results are relevant to real QCD as long as 
$m_s^2<4\mu\Delta$. In this case the order parameter
takes the form
\beq
\label{order_3}
\Delta^{ab}_{ij} = 
 \Delta (\delta_i^a\delta_j^b-\delta_i^b\delta_j^a),
 \hspace{0.5cm} (N_c=N_f=3).
\eeq
This is the color-flavor locked phase suggested in \cite{ARW_98b}. 
Both color and flavor symmetry are completely broken, but there 
are eight combinations of color and flavor symmetries that generate 
unbroken global symmetries. This is obvious from the result
(\ref{order_3}). We can always undo a flavor rotation by an
$SU(3)$ matrix $U$ by performing a subsequent color rotation
with the matrix $U^*$. Note that since color is a vector-like
symmetry, the remaining flavor symmetry is vector-like also. 
The symmetry breaking pattern is 
\beq
\label{sym_3}
SU(3)_L\times SU(3)_R\times U(1)_V\to SU(3)_V .
\eeq
This is exactly the same symmetry breaking that QCD exhibits 
at low density. This opens the interesting possibility
that in QCD with three flavors, the low and high density 
phases might be continuously connected \cite{SW_98b}. We
will discuss this phase in more detail in section 
\ref{sec_cfl}. 

 The phase structure of QCD-like theories with more than 
three flavors was studied in \cite{Sch_99}. The main result
is that the general structure found in the case of three
flavors extends to more than three flavors. The order 
parameter involves a coupling between color and flavor
degrees of freedom and the chiral $SU(N_f)_L\times 
SU(N_f)_R$ symmetry is broken down to a vector-like
subgroup. In general, however, this subgroup is not
the full flavor $SU(N_f)_V$ flavor symmetry that is 
present at zero density. In the case of four flavors,
for example, one finds $SU(4)_L\times SU(4)_R\to
O(4)_V$.
 
\section{Instantons and Color Superconductivity}
\label{sec_inst}

  The first quantitative estimate of the superconducting 
gap was given by Bailin and Love \cite{BL_84}. They assumed 
that the gap is generated by perturbative one-gluon 
exchange. In this case one has to deal with the problem 
that the quark-quark scattering amplitude is singular
at small angles. Bailin and Love assumed that in a 
cold quark liquid both the electric and magnetic 
interactions are screened at some scale $\Lambda$. 
In this case, the gap is on the order of $\Delta
\sim \mu\exp(-6\pi^2/(g^2L))$, where $L=\log(\mu^2/
\Lambda^2)$ depends logarithmicly on the screening
scale. This result implies that $\Delta$ is quite
small, unless we are willing to consider very large
coupling constants, $\alpha_s=g^2/(4\pi)>1$. 
Bailin and Love concluded that, most likely, 
$\Delta/\mu\sim O(10^{-3})$ for chemical potentials
$\mu<1$ GeV. We will see in the next section that 
the correct result for the perturbative gap 
is $\Delta\sim\mu g^{-5}\exp(-3\pi^2/(\sqrt{2}g))$,
which is parametrically bigger than the result
of Bailin and Love. Nevertheless, it remains unclear 
whether perturbation theory is applicable for chemical
potentials that are of practical interest. 
 
 After the work of Bailin and Love, there was practically
no interest in quark superconductivity until two groups
independently pointed out that non-perturbative effects,
instantons, can generate gaps that are much larger,
on the order of 100 MeV at densities a few times 
nuclear matter density \cite{RSSV_98,ARW_98}. In 
general it was realized that any kind of effective 
quark interaction which is consistent with the 
magnitude of the chiral condensate and general features
of the hadron spectrum at zero density will lead to 
gaps in that range. As we saw above, this is most
obvious in the case of two colors, where the diquark
condensate is related to the chiral condensate by
a symmetry. 

 Instantons are classical solutions of the euclidean
Yang-Mills field equations, describing tunneling 
events between topologically different vacuum 
configurations. Instantons are the strongest
non-perturbative fluctuations in the QCD vacuum; 
when lattice gauge configurations are made 
somewhat more smooth, it is instantons one 
mainly finds, see \cite{SS_98} for a recent review. 
Instantons have a dramatic effect on quarks. In
the background field of an instanton, the Dirac
operator has an exact, chiral, zero mode. This
zero mode is connected with the chiral anomaly, 
but it also plays an important role in connection
with spontaneous chiral symmetry breaking. There 
is evidence that the quark condensate can be 
understood as a collective state built from 
instanton zero modes. These zero modes, just 
like the chiral anomaly, continue to exist at
non-zero chemical potential. In this case, the 
zero modes lead to a pairing interaction, and
superconductivity. 

\begin{figure}[t]
\epsfxsize=12cm
\epsffile{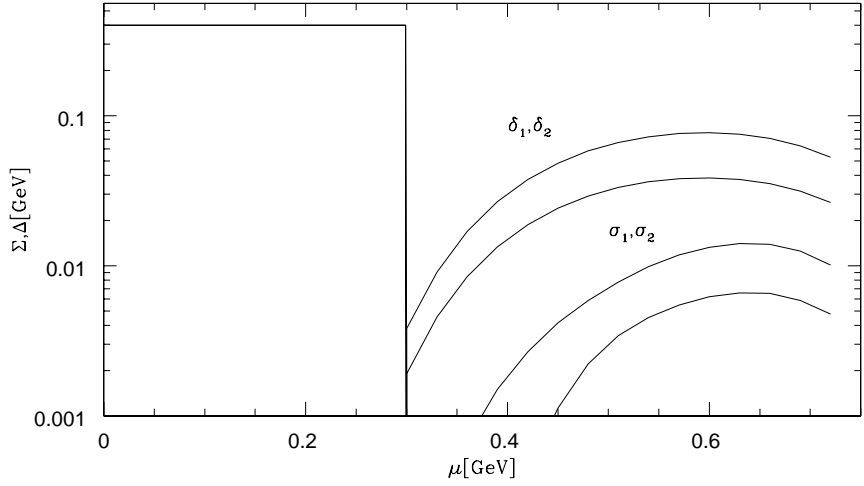}
\caption{\label{fig_inst}
Quark constituent mass and superconducting gap from a 
mean field calculation in $N_f=3$ QCD. }
\end{figure}

 This is most easily discussed in terms of effective
interactions. In the case of two flavors, the effect of
zero modes on the propagation of quarks can be encoded 
in a four fermion interaction
\beqa
\label{l_nf2}
{\cal L} &=& g\frac{1}{4(N_c^2-1)}
 \Big\{ \frac{2N_c-1}{2N_c}\left[
  (\bar\psi {\cal F}^\dagger\tau_\alpha^{-}{\cal F}\psi)^2 +
  (\bar\psi {\cal F}^\dagger\gamma_5\tau_\alpha^{-}{\cal F}\psi)^2
 \right] \cr
 & & \hspace{2cm}
  + \frac{1}{4N_c}(\bar\psi{\cal F}^\dagger \sigma_{\mu\nu}
\tau_\alpha^{-}{\cal F}\psi)^2 
\Big\} \ ,
\eeqa
where $N_c$ is the number of colors, $\tau^-=(\vec\tau,i)$ 
is an isospin matrix, and ${\cal F}$ is a form factor that
reflects the zero mode wave function \cite{CD_98,RSSV_99}.
From this interaction we can directly read off the effects
of instantons in different quark-anti-quark channels. In 
the pseudoscalar channel the interaction combines attraction 
for isospin 1 (pion) channel with repulsion for isospin 0 
($\eta'$). Similarly, one finds attraction in the scalar 
isospin-0 ($\sigma$) channel (responsible for spontaneous 
chiral symmetry breaking) together with repulsion in the 
scalar isospin-1 channel ($a_0$).  

  Using simple Fierz identities, we can construct the 
effective interaction for color antisymmetric $[\bar 3]$ 
and symmetric $[6]$ diquarks. The result is 
\beqa 
\label{l_diq}
{\cal L}_{diq} &=& 
{g\over 8 N_c^2} \left\{
 -{1\over N_c-1}
 \left[ (\psi^T{\cal F}^T C \tau_2 \lambda_A^a {\cal F}\psi)
        (\bar\psi{\cal F}^\dagger\tau_2 \lambda_A^a C {\cal F}^*\bar\psi^T  
\right. \right.\nonumber\\
 & &  \left.\left.\hspace{2cm} \mbox{}
       +(\psi^T{\cal F}^T C \tau_2 \lambda_A^a \gamma_5 {\cal F}\psi)
         (\bar\psi{\cal F}^\dagger \tau_2 \lambda_A^a \gamma_5 C {\cal F}^*
\bar\psi^T) \right]\right. \nonumber\\
 & & \left. \hspace{0.3cm}\mbox{}
      +{1\over 2(N_c+1)}
      (\psi^T{\cal F}^T C \tau_2 \lambda_S^a \sigma_{\mu \nu} {\cal F}\psi)
      (\bar\psi{\cal F}^\dagger \tau_2 \lambda_S^a \sigma_{\mu \nu} C 
{\cal F}^*\bar\psi^T) 
        \right\}  \  ,
\eeqa 
where $\tau_2$ is the anti-symmetric Pauli matrix, and $\lambda_{A,S}$ 
are the anti-symmetric and symmetric color generators. In the color 
$[\bar3]$ channel, the interaction is attractive for scalar 
$(\psi^TC\gamma_5\psi)$ diquarks, and repulsive for pseudoscalar 
$(\psi^TC\psi)$ diquarks. 

  We can now use this interaction in order to study the phase 
structure of $N_f=2$ matter \cite{BR_98,CD_98,RSSV_99}. This
is most easily done in the mean field approximation. At zero 
chemical potential chiral condensation is favored over diquark 
condensation, simply because the corresponding coupling constant 
is larger. At finite chemical potential chiral condensation is 
suppressed because quark loops are Pauli blocked. Quark-quark 
pairing, on the other hand, is enhanced because the density of 
states on the Fermi surface grows. As a result we find a first 
order transition from a phase with chiral symmetry breaking to a 
superconducting phase. As an example, we show the results of a 
somewhat more sophisticated calculation \cite{RSSV_99} in $N_f=3$ 
QCD in figure \ref{fig_inst}. The instanton density was fixed in 
order to reproduce a constituent quark mass $M=400$ MeV at zero 
chemical potential. In this case, we find the transition to the 
superconducting phase at a quark chemical potential $\mu_c\simeq 300$
MeV. In QCD with three flavors there is a flavor singlet and a flavor 
octet gap in the high density phase. The singlet gap peaks at around 
80 MeV. As we will discuss in more detail below, chiral symmetry remains 
broken in the high density phase. There are two types of constituent 
quark masses in the high density phase, but both of them are small, 
$M<10$ MeV. We should stress that the mean field description provides 
a very simple minded picture of the phase transition. The critical 
chemical potential is very small, and there is no nuclear matter phase 
in which quarks are clustered into nucleons.

\section{Superconductivity from perturbative one-gluon exchange}
\label{sec_oge}

If the chemical potential is very large, asymptotic
freedom implies that the coupling is weak. In this 
case, the gap can be calculated in weak coupling 
perturbation theory. The result $\Delta\sim\mu g^{-5}
\exp(-3\pi^2/(\sqrt{2}g))$ mentioned above was first
derived by Son using renormalization group arguments. 
It has since then been rederived in a number of works
\cite{SW_99b,PR_99b,HMSW_99,BLR_99}. Here we will 
follow \cite{SW_99b} and concentrate on the $N_f=2$
order parameter (\ref{order_2}). The gap matrix has
the form
\be
\label{gap_ans}
 \Delta^{ab}_{ij}(q) = (\lambda_2)^{ab}(\tau_2)_{ij}
 C\gamma_5 \left( 
     \Delta_1(q_0)\frac{1}{2}(1+\vec\alpha\cdot\hat q )
   + \Delta_2(q_0)\frac{1}{2}(1-\vec\alpha\cdot\hat q ) \right).
\ee
This decomposition is useful in order to discuss the weak
coupling limit and make the gauge invariance of the result
manifest. In weak coupling, we can replace $\vec\alpha\cdot\hat q
\to 1$ and only $\Delta_1$ produces a gap on the Fermi surface.
We have kept the dependence of the gap on the frequency. We
will justify the necessity for this below. In condensed
matter physics, keeping retardation effects in the gap 
equation is referred to as Eliashberg theory.

 We calculate the gap by solving the Dyson-Schwinger equation
for the quark self-energy in the Nambu-Gorkov formalism
\cite{BL_84}
\be
\label{ds}
 \Sigma(k) = -ig^2 \int \frac{d^4q}{(2\pi)^4}
  \Gamma_\mu^a S(q)\Gamma_\nu^b D^{ab}_{\mu\nu}(q-k).
\ee
Here, $\Sigma(k)=-(S^{-1}(k)-S_0^{-1}(k))$ is the proper 
self energy, $\Gamma^a_\mu$ is the quark-gluon vertex
and $D^{ab}_{\mu\nu}(q-k)$ is the gluon propagator. The
off-diagonal components of (\ref{ds}) determine the gap,
while the diagonal components fix the wave function 
renormalization. Since the fermion momenta are large 
we will ignore wave function renormalization. For the 
same reason, we can also ignore vertex corrections. 

 The gluon momentum, however, can become soft and 
corrections to the gluon propagator have to be taken
into account. The gluon propagator in a general 
covariant gauge is
\be
\label{D_dec}
 D_{\mu\nu}(q) = \frac{P_{\mu\nu}^T}{q^2-G} 
 + \frac{P_{\mu\nu}^L}{q^2-F} - \xi\frac{q_\mu q_\nu}{q^4},
\ee
where $P_{\mu\nu}^{T,L}$ are transverse and longitudinal 
projectors and $G,F$ are functions of energy and momentum.
$\xi$ is a gauge parameter that should not appear in any 
physical result. 

 We can derive two coupled equations for the gap parameters
$\Delta_{1,2}(p_0)$. In the weak coupling limit, we find that
only terms involving $\Delta_1$ have a singularity on the 
Fermi surface. We can therefore drop $\Delta_2$. The resulting
equation for $\Delta_1$ is independent of the gauge parameter. 
We find
\bea
\label{gap_5}
\Delta(p_0) &=& \frac{g^2}{12\pi^2} \int dq_0\int d\cos\theta\,
 \left(\frac{\frac{3}{2}-\frac{1}{2}\cos\theta}
            {1-\cos\theta+(G+(p_0-q_0)^2)/(2\mu^2)}\right. \\
 & & \hspace{3cm}\left.    +\frac{\frac{1}{2}+\frac{1}{2}\cos\theta}
            {1-\cos\theta+(F+(p_0-q_0)^2)/(2\mu^2)} \right)
 \frac{\Delta(q_0)}{\sqrt{q_0^2+\Delta(q_0)^2}}. \nonumber
\eea
The integral over $\cos\theta$ is dominated by small $\theta$, 
corresponding to almost collinear scattering. It is therefore 
important to take medium modifications of the gluon propagator
at small momenta into account. For $\vec{q}\to 0$ and to 
leading order in perturbation theory we have
\be
 F = 2m^2, \hspace{1cm}
 G = \frac{\pi}{2}m^2\frac{q_0}{|\vec{q}|},
\ee
with $m^2=N_fg^2\mu^2/(4\pi^2)$. In the longitudinal part,
$m_D^2=2m^2$ is the familiar Debye screening mass. In the 
transverse part, there is no screening of static modes, 
but non-static modes are modes are dynamically screened
due to Landau damping. In our case, typical frequencies
are on the order of the gap, $q_0\simeq \Delta$. This means
that the electric part of the interaction is screened at
$q_E\simeq m_D^{1/2}$ whereas the magnetic interaction
is screened at $q_M\simeq (\pi/4\cdot m_D^2\Delta)^{1/3}$.
Asymptotically, $q_M\ll q_E$, and magnetic gluon exchange
dominates over electric gluon exchange. 

 We can now perform the angular integral and find
\be
\label{eliash_mel}
\Delta(p_0) = \frac{g^2}{18\pi^2} \int dq_0
 \log\left(\frac{b\mu}{|p_0-q_0|}\right)
    \frac{\Delta(q_0)}{\sqrt{q_0^2+\Delta(q_0)^2}},
\ee
with $b=256\pi^4(2/N_f)^{5/2}g^{-5}$. We can now see why 
it was important to keep the frequency dependence of the 
gap. Because the collinear divergence is regulated by
dynamic screening, the gap equation depends on $p_0$
even if the frequency is small. We can also understand
why the gap scales as $\exp(-c/g)$. The collinear
divergence leads to a gap equation with a double-log
behavior. Qualitatively
\be
\label{dlog}
 1 \sim \frac{g^2}{18\pi^2}
 \left[\log\left(\frac{\mu}{\Delta}\right)\right]^2,
\ee
from which we conclude that $\Delta\sim\exp(-c/g)$. 
The approximation (\ref{dlog}) is not sufficiently
accurate to determine the correct value of the 
constant $c$. For this purpose, we have to solve
the integral equation (\ref{eliash_mel}). This 
can be done by converting the integral equation
into a differential equation \cite{Son_98}. In 
the weak coupling limit, an approximate solution 
is given by
\be
\label{sol_son}
\Delta(p_0) \simeq \Delta_0 \sin\left(\frac{g}{3\sqrt{2}\pi}
 \log\left(\frac{b\mu}{p_0}\right)\right),\hspace{0.5cm}
 p_0>\Delta_0,
\ee
with $\Delta_0=b\mu\exp(-3\pi^2/(\sqrt{2}g))$. 
The final result for the magnitude of the gap on the Fermi 
surface is 
\beq
\label{gap_oge}
\Delta_0 \simeq 256\pi^4(2/N_f)^{5/2}\mu g^{-5}
   \exp\left(-\frac{3\pi^2}{\sqrt{2}g}\right).
\eeq
We should emphasize that, strictly speaking, this result
contains only an estimate of the pre-exponent. This estimate
is obtained by collecting the leading logarithms from 
electric and magnetic gluon exchanges. We also note that
the result shows that the perturbative calculation is 
self-consistent. Asymptotically, the factor $\mu$ in 
the pre-exponent overwhelms the exponential suppression
factor $\exp(-c/g)$ and $\Delta\gg\Lambda_{QCD}$. This
implies that the magnetic screening scale $q_M\sim 
(g^2\mu^2\Delta)^{1/3}\gg\Lambda_{QCD}$ and the
perturbative result for the gluon polarization
function is reliable. 

 For chemical potentials $\mu<1$ GeV, the coupling 
constant is not small and the applicability of perturbation
theory is in doubt. It is known that at finite temperature
the convergence properties of the perturbative expansion
are extremely poor. It has been argued that the situation
at non-zero chemical potential might be better \cite{PR_99b}.
The superconducting gap, however, is dominated by collinear
exchanges. For $\mu<1$ GeV we have $q_M\simeq \Lambda_{QCD}$
and the gap is mainly determined by very small momenta. If
we ignore this problem and calculate the gap at chemical
potentials on the order of 500 MeV, we find $\Delta\simeq
100$ MeV. It is gratifying to see that this result is quite
close to the low-density estimate based on instantons or
other effective interactions.

\section{Color-Flavor Locking and Quark-Hadron duality
in QCD with three flavors}
\label{sec_cfl}

 In this section we wish to study the color-flavor locked
phase in QCD with $N_f=3$ flavors in somewhat more detail.
In the limit where all quarks are massless the underlying 
theory has the continuous symmetry group $SU(3)_C \times 
SU(3)_L \times SU(3)_R \times U(1)_B$. In the color-flavor
locked phase the symmetry is broken following the pattern
\be
SU(3)_C \times SU(3)_L \times SU(3)_R \times U(1)_B \rightarrow
SU(3)_{C+L+R} \times Z_2. 
\ee
The breaking of local color symmetry implies that all the gluons
acquire mass, according to the Higgs effect. There are no long-range 
interactions.  At low energy, there is no direct signature for the 
color degree of freedom. Color is, in this sense, confined. 

 As explained in section \ref{sec_nf}, chiral symmetry is 
broken because color, being a vector-like interaction, locks
left handed flavor rotation to right handed flavor rotations.
This is an unusual mechanism for chiral symmetry breaking.
The condensates $\langle\psi_L\psi_L\rangle$ and $\langle
\psi_R\psi_R\rangle$ of left and right handed quarks are quite
separate. Indeed, unless we include instantons, they are not
even phase coherent. Nevertheless, because both are locked
to color they are thereby locked to one another.

 Superconductivity breaks the $U(1)$ of baryon number. This
might seem disconcerting at first, but the total baryon number 
of a finite sample is still conserved. The meaning of baryon
number violation is that there is a Goldstone mode, the 
phonon, that corresponds to small fluctuations in the local
baryon density. The macroscopic consequence of this fact is 
easy transport of baryon number, or superfluidity. We can
construct a gauge invariant order parameter for superfluidity
by convoluting several of the primary, diquark, condensates:
\be 
 \chi = \langle \epsilon^{abc}\epsilon_{ijk}
    \phi^a_i\phi^b_j\phi^c_k \rangle,   \hspace{1cm}
 \phi^a_i=\epsilon^{abc}\epsilon_{ijk}\psi^b_j C\gamma_5\psi^c_k .
\ee
We note that this order parameter carries the quantum 
numbers of a di-lambda, or the hypothetical H-dibaryon.

 A diquark condensate breaks baryon number but there is
a residual $Z_2$ symmetry that corresponds to flipping the sign
of the quark fields. The primary condensate also leaves an 
axial $Z_2$ symmetry, but in QCD this symmetry is broken 
by instantons. At large chemical potential instantons and 
other semi-classical fluctuations are suppressed, and the 
axial $Z_2$ symmetry is almost exact. One consequence of this
result is that the ordinary chiral condensate $\langle \bar
\psi\psi\rangle$ is strongly suppressed, whereas its 
square $\langle(\bar\psi\psi)^2\rangle$ is not small, 
proportional to the primary diquark condensate squared
\cite{Sch_99}. 

 It is interesting to consider how the photon couples to
color-flavor locked matter. The original electromagnetic gauge 
invariance is broken, but there is a combination of the original 
electromagnetic gauge symmetry and a color transformation which 
leaves the condensate invariant. The original photon $\gamma$
couples with strength $e$ to the flavor matrix ${\rm diag}(2/3,
-1/3,-1/3)$. There is a diagonal gluon $G$ which couples with 
strength $g$ to the color matrix ${\rm diag}(-2/3,1/3,1/3)$.
Then the combination 
\be
\label{new_gamma}
\tilde \gamma = \frac{g\gamma+eG}{\sqrt{e^2+g^2}}
\ee
leaves the color-flavor locked condensate invariant. In the
color superconducting medium it represents the physical photon.  
This photon couples to electrons with coupling $eg/\sqrt{e^2+g^2}$,
which is the unit of charge in the color-flavor locked phase. 
The coupling to quarks receives contributions from both the 
color and the flavor charges. It is easy to see that, in 
units of the electron charge, these contributions always
add up to $\pm 1$ or 0. The same is true for gluons, too.

 We are now in a position to examine the low energy excitations
in the color-flavor locked phase. The spontaneous breaking of 
global chiral $SU(3)_L \times SU(3)_R$ brings with it an octet 
of pseudoscalar Nambu-Goldstone bosons. In addition to that 
there is a massless phonon, corresponding to the spontaneous
breaking of $U(1)_B$. If the density is very high, there is
also an anomalously light pseudo-Goldstone boson associated
with the $U(1)_A$ symmetry.

 In addition to the Goldstone modes, there are also excitations
directly derived from the original quark and gluon modes. The
quarks form an octet and a singlet of spin 1/2 particles under 
the residual $SU(3)_V$. There is an energy gap for the production 
of quark pairs. Dynamical calculations show that this gap is larger 
for the singlet than for the octet. This means that the states
in the octet are true long-lived quasiparticles, since there is 
nothing for them to decay into. We can think of these states
as baryons, since baryon number is only conserved modulo 2/3.
The gluon fields form an octet of spin $1$ bosons under the 
$SU(3)_V$ symmetry. They acquire a mass by the Meissner-Higgs 
phenomenon. 

   The universal features of the color-flavor locked state: 
confinement, chiral symmetry breaking down to vector $SU(3)$, 
and superfluidity, are just what one would expect for $N_f=3$ 
QCD at low density. We have now seen that the same is true for 
the spectrum of low lying non-Goldstone excitations: There is 
an octet and singlet of baryons, and an octet of vector mesons. 
This means that there is no sharp distinction between the 
low-density, hadronic, phase and the high-density, quark, phase. 
In particular, there need not be a phase transition separating 
the high and low density phases of $N_f=3$ QCD. Furthermore, 
we have realized and old goal of QCD: we have obtained a weak 
coupling, but non-perturbative, description of hadronic degrees 
of freedom in terms of quarks and gluons.

\section{Chiral density waves}
\label{sec_cdw}

 Throughout this contribution we have assumed that the dominant 
instability at large density corresponds to the formation of
particle-particle (or hole-hole) pairs. This is a very reasonable
assumption, because only this kind of pairing occurs for arbitrarily 
weak coupling and uses the whole Fermi surface coherently. Nevertheless, 
in strong coupling, or if superconductivity is suppressed, other forms 
of pairing may take place. Obvious candidates are the formation of 
larger clusters (nucleons, four-quark states, dibaryons) or 
particle-hole pairing \cite{DGR_92,SS_99,PRWZ_99}.

  Particle-hole pairing is characterized by an order parameter of
the form
\be 
\langle\bar\psi(x)\psi(y) \rangle = 
 \exp(i\vec{p}\cdot(\vec{x}+\vec{y}))\Sigma(x-y),
\ee
where $|\vec{p}|=p_F$ is a vector on the Fermi surface. This 
state describes a chiral density wave. It was first suggested 
in \cite{DGR_92} as the ground state of QCD at large chemical
potential and large $N_c$. The color factors for particle-particle
and particle-hole scattering are
\be
c=\frac{N_c+1}{2N_c}   \hspace{0.5cm} (pp),\hspace{1cm}
c=\frac{N_c^2-1}{2N_c} \hspace{0.5cm} (ph),
\ee
suggesting that particle-particle pairing, and superconductivity,
is suppressed at large $N_c$. Particle-hole pairing is not 
suppressed, but suffers from the fact that a direction is 
singled out and only a small part of the Fermi surface is
used. At large $N_c$, however, screening due to fermions 
is weak and the perturbative one-gluon exchange interaction 
is even more strongly dominated by collinear exchanges.
This favors the particle-hole instability. 

\begin{figure}[t]
\epsfxsize=12cm
\epsffile{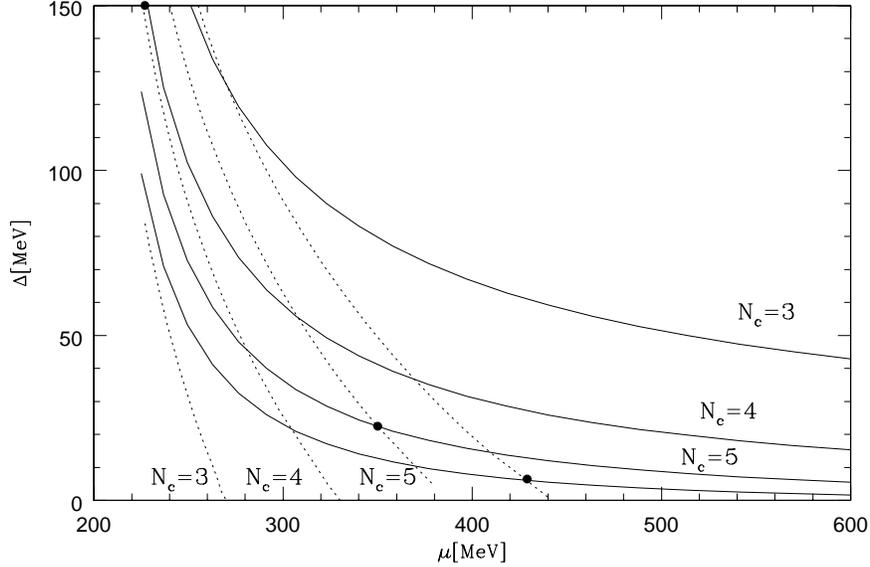}
\caption{\label{fig_cdw}
BCS (solid line) and chiral density wave (dashed line) gaps from 
a perturbative calculation in QCD with $N_c$ colors and $N_f=2$ 
flavors. }
\end{figure}
 
  Screening was ignored in the original work of \cite{DGR_92},
but is taken into account in the more recent investigations
\cite{SS_99,PRWZ_99}. The main conclusion is that, in weak
coupling, the chiral density wave instability requires very
large $N_c\gg 3$. We have verified this fact from numerical
solutions of the chiral density wave gap equation. In 
perturbative QCD, the gap equation reads \cite{PRWZ_99}
\beqa
\Sigma(p_{||}) &=& c\,\frac{g^2}{12\pi^2}\int dq_{||} 
\left\{\log\left( 1 +\frac{\Lambda_{\perp}^3}
         {(p_{||}-q_{||})^3+\frac{\pi}{4}m_D^2|p_{||}-q_{||}|}
           \right) \right. \nonumber \\
 & & \hspace{2cm}\left. 
 + \frac{3}{2}\log\left( 1 +\frac{\Lambda_{\perp}^2}
         {(p_{||}-q_{||})^2+m_D^2}\right)\right\}
        \frac{\Sigma(q_{||}}{\sqrt{q_{||}^2+\Sigma(q_{||})^2}}.
\eeqa
This gap equation is very similar to the BCS gap equation
(\ref{eliash_mel}), except that the color factor is different,
and that the angular integration is cut off at a scale
$\Lambda_\perp=\sqrt{2\mu q_{||}}$. This fact reflects the
one-dimensional nature of the chiral density wave. 
 
  Numerical results for both the BCS and chiral density 
wave (CDW) gap are shown in figure \ref{fig_cdw}. We find
that for moderate $N_c$, in particular in the case $N_c=3$,
the chiral density wave instability only exists for very
small $\mu$ and very large coupling. In the case of $N_c=3$,
the CDW gap is never bigger than the BCS gap, even if the 
coupling is big. 

  These results are not really quantitative. Only for very
large $N_c$ does the chiral density wave instability occur 
in weak coupling where the perturbative analysis is 
appropriate. Nevertheless, our results illustrate the 
point that in the real world, with $N_c=3$, superconductivity
is the dominant instability even at moderate chemical 
potential. Using more realistic interactions in the 
strong coupling regime is not likely to change this 
situation, since other interactions, like instantons, 
will typically lack the strong collinear enhancement of 
the one gluon exchange interaction. In addition to that,
the true ground state is not determined by the size of
the gap, but by the magnitude of the condensation energy. 
Since the chiral density wave state does not use the
whole Fermi surface, the condensation energy is only
of the order $\mu\Sigma^3$, as compared to $\mu^2
\Delta^2$ in the case of superconductivity.

\section{Conclusion: The many phases of QCD}

\begin{figure}[t]
\epsfxsize=12cm 
\epsfbox{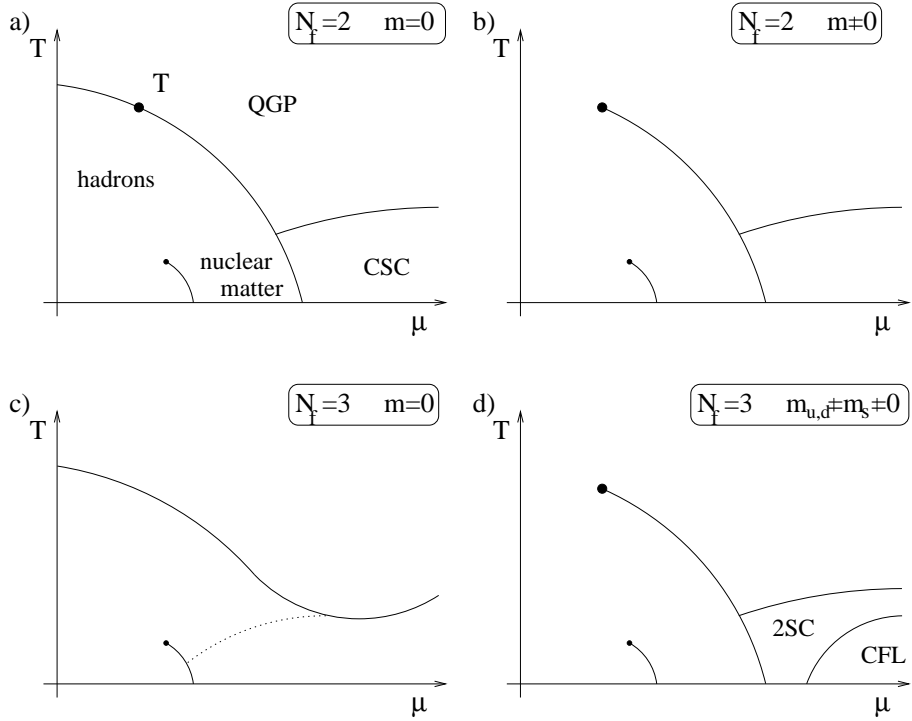}
\caption{Schematic phase diagram of QCD at finite temperature
and density. The figures a)-d) correspond to different numbers
of massless and massive flavors, see the discussion in the text.  
\label{fig_phase}}
\end{figure}

 We would like to conclude by summarizing some of the things
we have learned about the phase structure of QCD-like theories
at finite temperature and chemical potential. We begin with the 
case of two massless flavors, figure \ref{fig_phase}a. If we
move along the chemical potential axis at temperature $T=0$, there
is a minimum chemical potential required in order to introduce 
baryons into the system. Since nuclear matter is self-bound,
this point is a first order transition: The density jumps from
zero to nuclear matter density. Along the temperature
axis, the line of first order transitions eventually ends in 
a critical point: This is the endpoint of the nuclear liquid-gas
phase transition. If we continue to increase the chemical potential,
we encounter the various phases of nuclear matter at high density. 
Many possibilities have been discussed in the literature, and 
we have nothing to add to this discussion. At even higher 
chemical potential, we encounter the transition to quark matter
and the two flavor quark superconductor. Model calculations
suggest that this transition is first order. This is also
consistent with the fact that, asymptotically, the 
superconductor is of the first kind, and expected to
exhibit a first order transition to the normal phase at
the critical temperature for color superconductivity.
In the case of two massless flavors, universality 
arguments suggest, and lattice calculations support,
the idea that the finite temperature zero chemical 
potential chiral phase transition is second order. In
this case, the line of first order $\mu\neq 0$ transitions
meets the $T\neq 0$ transition at a tricritical point 
\cite{BR_98,HJS*_98}. 

  This tricritical point is quite remarkable, because it 
remains a true critical point, even if the quark masses are
not zero, figure \ref{fig_phase}b. A non-zero quark mass
turns the second order $T\neq 0$ transition into a smooth
crossover, but the first order $\mu\neq 0$ transition
persists. While it is hard to predict where exactly 
the tricritical point is located in the phase diagram it
may well be possible to settle the question experimentally. 
Heavy ion collisions at relativistic energies produce 
matter under the right conditions and experimental 
signatures of the tricritical point have been suggested
\cite{SRS_98}.

  We have already discussed the phase structure of $N_f=3$
QCD with massless or light degenerate quarks in section
\ref{sec_cfl}. We emphasized that at $T=0$ the low density,
hadronic, phase and the high density, quark, phase might
be continuously connected. On the other hand, there has
to be a phase transition that separates the color-flavor
locked phase from the $T=\mu=0$ hadronic phase. This is
because of the presence of a gauge invariant $U(1)$ order
parameter that distinguishes the two. In the case of 
$N_f=3$ massless flavors the finite temperature phase
transition is known to be first order. We expect the 
transition from the superconducting to the normal phase
at $T\neq 0$ and large $\mu$ to be first order, too. This
means that there is no tricritical point in figure
\ref{fig_phase}c. 

  The phase diagram becomes more complicated if we take
into account the effects of a finite strange quark mass,
figure \ref{fig_phase}d. Consider increasing the strange
quark mass in the color-flavor locked.phase. This shifts
the Fermi momentum of the strange quarks with respect to
the light quarks. If $m_s^2/(4\mu)$ is larger than $\Delta
(m_s\!=\!0)$ pairing between strange quarks and light
quarks can no longer take place, and there is a first order
transition to a phase (2SC) with separate pairing in the 
$ud$ and $s$ sectors \cite{SW_99,ABR_99}. Asymptotically
the gap grows with $\mu$ and we expect the color-flavor 
locked phase to dominate for any value of $m_s$. Increasing
the temperature or reducing the chemical potential favors 
the 2SC phase. Whether in the case of the physical value 
of the strange quark mass there is an interlude of the 
the 2SC phase along the $\mu\neq 0$ axis, instead of a 
direct transition between the CFL phase and nuclear matter, 
current calculations are not sufficiently accurate to decide. 
We know that there is at least one phase transition, because 
nuclear matter and the color-flavor locked phase are distinguished 
by a gauge invariant $U(1)_s$ order parameter. This, of 
course, is based on our belief that nuclear matter is
stable with respect to strange quark matter or hyperonic
matter. Current calculations have also not conclusively
answered the question whether the transition along the
$T\neq 0$ axis is a smooth crossover, as indicated in 
figure \ref{fig_phase}d and favored by some lattice 
calculations, or whether the transition is first order,
as would be the case if $m_s$ is sufficiently small. 
This is clearly an important question in connection 
with the existence of the tricritical point. 

  Thinking about QCD at finite baryon density has 
taught us many interesting lessons about the phase
diagram, but there remain many question marks in 
the picture presented here, and much work remains
to be done.

\end{document}